# Rings in the Solar System: a short review

Sébastien CHARNOZ[1], Aurélien CRIDA[2,3], Ryuki HYODO[4]

(1) Université Paris Diderot / Institut de Physique du Globe, Paris FRANCE
(2) Université Côte d'Azur / Observatoire de la Côte d'Azur, Lagrange, Nice, FRANCE
(3) Institut Universitaire de France, Paris, FRANCE
(4) Earth-Life Science Institute/Tokyo Institute of Technology, 2-12-1 Tokyo, Japan

**Abstract:**
**Rings are ubiquitous around giant planets in our Solar System. They evolve jointly with the nearby satellite system. They could form either during the giant planet formation process or much later, as a result of large scale dynamical instabilities either in the local satellite system, or at the planetary scale. We review here the main characteristics of rings in our solar system, and discuss their main evolution processes and possible origin. We also discuss the recent discovery of rings around small bodies.**

All giant planets of the Solar System have rings. The brightest and most massive, those of Saturn, have been discovered by Galileo Galilei himself (whereas he did not interpreted them as rings, the first one to interpret them correctly was the dutch astronomer Christian Huygens), but it is really with the space missions Voyager 1 and 2, during the 70's and 80's that it was realized that all four giant planets harbor rings. Conversely, rings seem absent around terrestrial planets, despite many attempts to find dusty rings (especially around Mars).
Rings are still some very mysterious structures that are ubiquitous to all giant planets, either gas or ice giants. However, the 4 rings systems that we know are all very different, and within a ring system, rings could be either dense and made of large particles (like Saturn and Uranus rings) or dusty (like Jupiter or Saturn's E or G rings). See Figure 1 for a comparative sketch of the 4 rings systems found in our Solar System. Unlike a very common thought, rings are not necessarily confined to a planet's Roche limit (see section 2), and dusty rings are found commonly beyond the Roche Limit. The generality and the diversity of rings systems make their origin difficult to understand. They could form either concomitant with their host planets (inside the circumplanetary disk), or they could be the result of late dynamical evolution either of their satellite systems (through collisions), or results from comet showers due to large scale instabilities of the planetary system. Thanks to the Cassini and Galileo missions, Saturn and Jupiter rings were investigated in great details (see e.g. Cuzzi et al., 2010 for a review of Saturn's rings as seen by Cassini). Today, due to the increased detection capacities, Uranus and Neptune rings are better understood thanks to observations from Earth (see e.g. Showalter et al., 2008). Rings are rapidly evolving structures under the action of gravity, viscosity and radiative forces. We understand now that there is a close association between rings and satellites, through dynamical interactions (moons sculpt rings) or material exchange (rings can give birth to moons, see e.g. Charnoz et al., 2011; Crida & Charnoz 2012). As they represent a huge surface to volume ratio, rings are also very efficient to capture meteoritic bombardment, that, over long timescales, may modify their average composition. We have no clear idea of the ring's age. They could be as old as the Solar System itself (like in the formation scenario proposed by Canup 2010 invoking a migrating satellite that is tidally dirupted), or as young as a few 100 Myrs according to different theories (like in Ćuk et al., 2016). But no model is consistent with all data, and the planetary rings origin is still heavily debated. Finally, unexpectedly, rings were found recently about small bodies, like centaurs, making a general theory of their formation even more difficult to build.

In this chapter, we will briefly review our current knowledge about rings in our Solar System, emphasizing their diversity, as well as the key fundamental processes that govern their evolution. We also discuss the recent rings detection about small bodies.

# 1.Overview of Solar System rings systems

Main rings systems are those that are found inside the planet's Roche Limit (see section 2), i.e. typically inside ~2.5 planetary radii. Of course, the definition of the Roche Limit depends on the ring material density, that could be diverse and changing, and that is often not very well constrained. However, the Roche Limit only scales with the density to the exponent 1/3, so it does not vary greatly from one system to another and is found in general around 2-3 planetary radii. Outside the Roche Limit, dusty rings systems are also found, but they need to be replenished because the material is removed by the Poynting-Robertson force. In addition, the collisional timescale of dusty rings system is so long, that they are never flat and are in general vertically extended either due to radiative forces, or simply by keeping the memory of the orbit of the moon that is the dust source .

We now briefly present the main characteristics of the rings systems around the 4 giant planets.

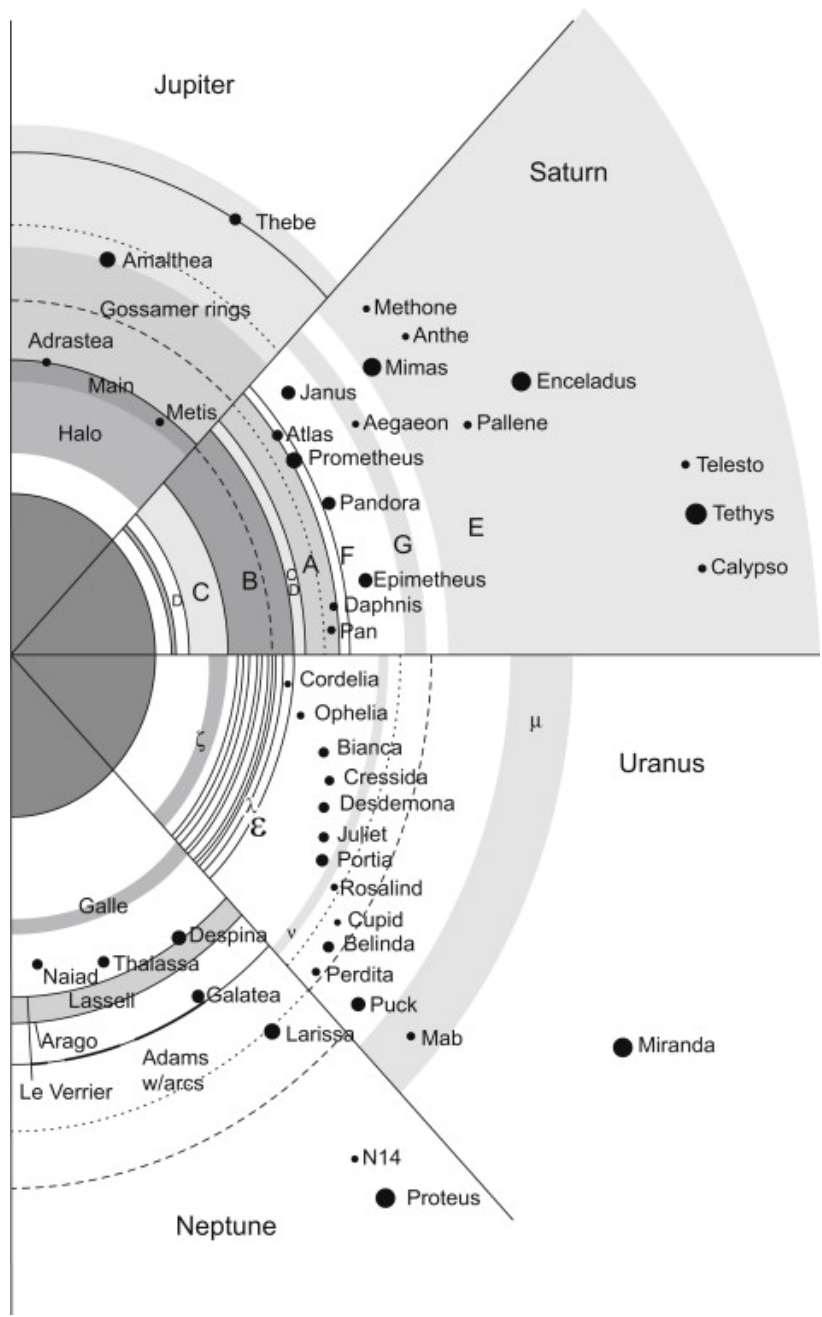

Figure 1: Jupiter, Saturn, Uranus and Neptune rings and inner moons scaled by the planet radius. The shaded regions designate the dusty rings.

**1.1 Jupiter rings**

Jupiter rings are extremely tenuous and are mostly dusty (Figure 2). They are closely associated to the four small moons, Methis, Adrastea, Amalthea and Thebe that release dust from their orbit. This dust adopts a 3D structure with finite thickness due to the randomization of their longitude of nodes and pericenters. The two outermost rings (the "Amalthea" and the "Thebe" gossamer rings), extend up to about 2.5 and 3.15 jovian radii respectively and have an optical depth about $10^{-7}$ and $10^{-8}$. They are vertically extended (unlike Saturn's main rings) with thickness about 2300 and 8400 km respectively, that reflects the inclination of the two parent moons. The volume density of these two rings is not maximum in the midplane, but rather at the top and the bottom, because of the simple geometric effect : a particle on an inclined orbit (like all dust grains in these rings) spend more time at the top and bottom of its orbit because the vertical velocity is there zero. This creates a "sandwich"-like structure whose origin is purely kinematic. Closer to the planet the main rings, bounded by Adrastea is much flatter, about 30 km and is composed mainly of big particles as testified by a weak transmission in forward scattered light. A vertical extension appears, up a to a few 100 km, that may be composed of micrometer sized dust. Then comes, the innermost ring, called the "halo" that is very puffed up (up to a few 10,000 km) and is dusty. The vertical extension of the halo is maybe due to the coupling of the charged dusty grains with the magnetic field of Jupiter, that excites the particles inclinations (Hamilton et al., 2008). Dust circulates efficiently in this ring system due to the Poynting-Robertson force (see section 2), forcing the dust grains to spiral toward the central planet.

Particle sizes in Jupiter's rings system show a steep size distribution, and are typically in the range 0.1-30 $\mu$m. Only the main ring seems to contain a significant population of cm sized particles. The structure of the Jovian rings is currently the best understood the four rings systems. It shows well that big particles, that are not subject to radiative forces, and lose energy through collisions gather in a thin sheet (the main rings) whereas dusty particles, subject to radiative forces and that are in a low collisional environment (owing to the lower optical depth in the dusty rings) in general do not flatten into a thin rings system, but keep a torus like structure in 3D.

For a detailed review of Jupiter rings and physics, we recommend the Burns et al. (1984) review paper.

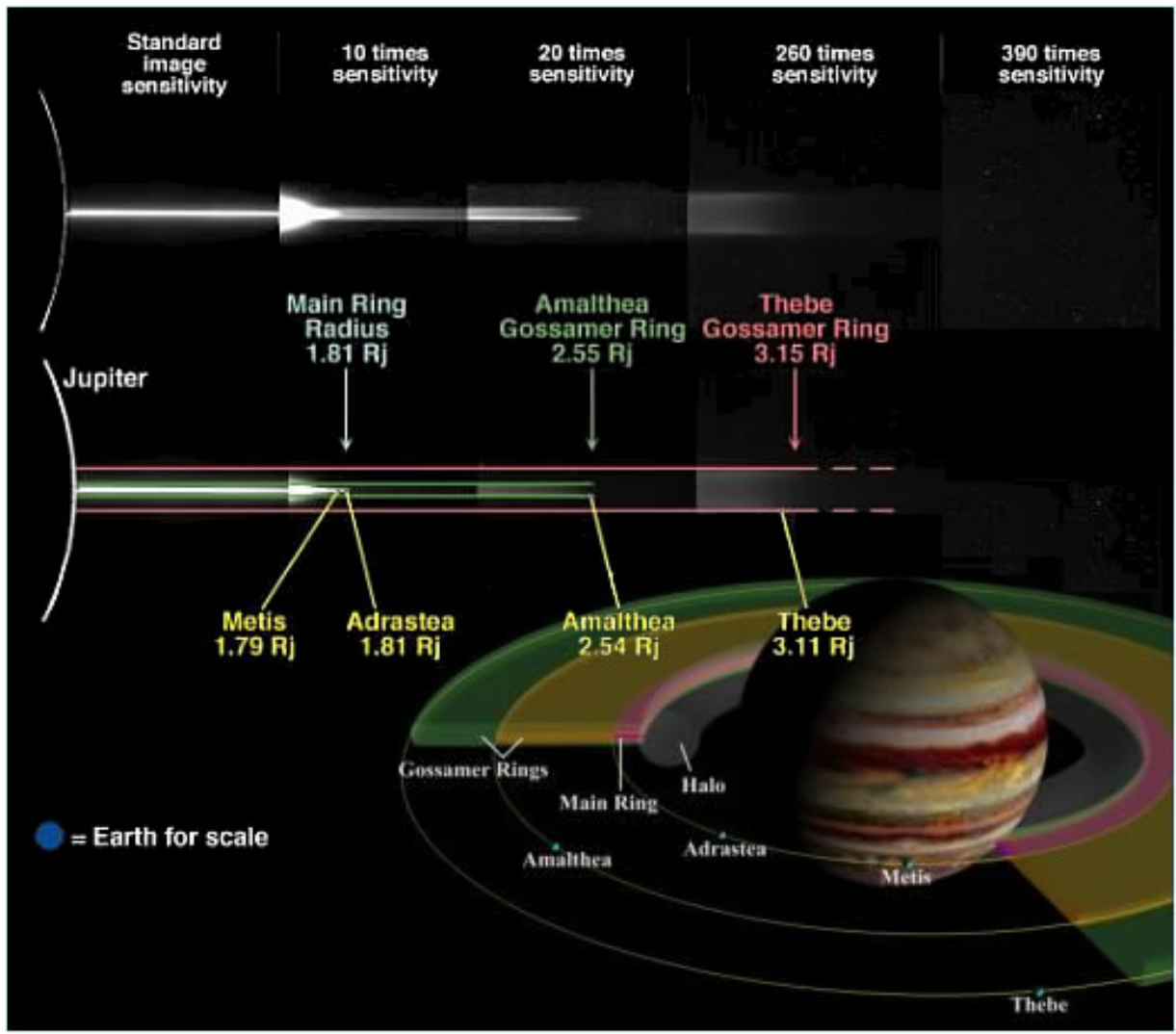

Figure 2 : Jupiter ring's structure. Rj=Jupiter' radius

**1.2 Saturn's rings**

They are the most famous, the richest and the most massive ring system of the four giant planets. Very broadly speaking, they are composed of a dense and vertically thin ring system inside the Saturn's Roche limit (about 2.5 Saturn's radii), as well as several of dusty rings, sometimes associated with satellites, beyond the Roche limit. Hundreds of dynamical structures have been identified in Saturn's main ring system, with a number of them clearly associated to ring-moons gravitational interactions, but the origin of numerous structures is not clearly associated to moons. The study of Saturn's rings is a very dynamical subject, fueled by the Cassini mission, and several review papers can be found in the "Saturn from Cassini-Huygens" book (2009, M. Dougherty, L.Esposito, S. Krimigis Eds. Springer) and the forthcoming "Planetary Rings" book (C.D. Murray, M. Tiscareno Eds, Univ. of Arizona Press), that should be edited in 2018. For a review of the most surprising structures we recommend the Cuzzi et al. (2010) paper or the (excellent) Tiscareno (2013) review paper about rings in the Solar System.

The main rings systems consist in 4 -regions with different optical depth and different types of structures (Figure 3). By increasing order of distance to Saturn, we find first the D and C rings, deeply embedded in Saturn's Roche limit, that consists of ringlets and plateaus of material with varying optical depth about $10^{-4}$ and $10^{-3}$ respectively, then comes the B ring, the densest ring and probably the one that contains most of Saturn's rings mass, with optical depth much higher than 1. Outside the B ring, we find the Cassini division, a low optical depth regions rich in ringlets and dusty plateaus, and then finally, we find the translucent A ring, with optical depth about 1, and that extends up to about 136,000 km. A thin, and very dynamical ringlet, the F ring, lays precisely at Saturn's Roche limit (about 140,000 km). It is bounded inside and outside by two moons, Prometheus and Pandora, whose complex dynamical interactions triggers a wide variety of structures in the F ring. Prometheus and Pandora are called "Shepherd" moons, because they were thought,once, to gravitationally confine the F ring. Transient accretion structures (like clumps) have been identified, but seems very ephemeral (Beurle et al., 2010).

Whereas no moon have been found in the D, C, B and the Cassini division, the A ring contains several moonlets that create gap and density waves (Pan, Daphnis, Atlas), but more distant moons also triggers waves (Prometheus, Pandora, Janus, Epimetheus etc.) . Making a list of all dynamical structures found in Saturn's main rings is out of the scope of the present chapter, but among the most notable structures we find numerous spiral density waves in the A ring and some in the B ring (see e.g. Tiscareno et al., 2007). The outer edge of Saturn's B ring shows complex vertical structures and seem to be maintained by the exchange of angular momentum with the Mimas moon, which is in 2:1 resonance with the B ring outer edge (Cuzzi et al., 2010, Spitale and Porco 2010). Inclined moons creates also vertical waves in the main rings (called "bending waves') visible in the A ring. The main ring's thickness has been estimated using different techniques (numerical simulations, study of heat conductions) and is estimated about 10 meters (see e.g. Reffet et al., 2015). Particles in the main rings are composed of water ice, with a slight contamination (less that 1% in mass) by some unknown red material that could be silicate or iron (Cuzzi et al., 2009). This peculiar composition (>99% of water ice) that is very different from the average composition of outer solar system objects. It is thought to be a very constraining a clue to their origin (see e.g. Charnoz et al., 2009, Canup 2010), as well as a challenge to be explained. The mass of Saturn's main ring is a matter of debate. Study of density waves in the A ring, give an idea of the local surface density. Extrapolation to the entire ring system gives a total mass about $10^{19}$ to $10^{20}$ kg, consistent with other independent estimates (see e.g. Esposito 2010, Reffet et al., 2015). Interestingly, with this mass, and assuming a thickness of a few meters, this implies that Saturn's ring is a self-gravitating disk, i.e., it may develop spiral structures patterns. Indeed, spiral patterns have been found, but at the km scales called "wakes" which is explained by the strong keplerian shear in Saturn's rings see Schmidt et al., (2009) and Cuzzi et al., (2010). This may have important implication for the ring long term evolution, as it is known (see e.g. Salmon et al., 2010) that self-gravitating astrophysical disk are self-regulated by a balance between their surface density and velocity dispersion and maintain their surface density close to a Toomre coefficient close to 1.

Star occultations allowed to derive size distribution in different regions of the ring system. Saturn's main rings do not contain dust, and particle sizes are in the range 1cm to a few 10 meters. The size distribution show variations depending on the region in the rings, with the largest size in general increasing with distance to Saturn, that may be a sign of more efficient accretion at large distance (Cuzzi et al., 2009). Dust (i.e micrometer sized grains) are in general absent from the main rings, except in a handful of very narrow ringlet like in the Encke gap or in the Laplace gap.

Beyond the main rings, dusty rings are also found, and have been detected in forward scattered light thanks to Voyager and Cassini missions. The G ring presents long arc structures (i.e. incomplete rings) maybe due to the erosion of an unseen population of km-sized moonlets, but still undetected. Saturn's E ring, is a dusty rings and results from the ejection of icy dust in Enceladus' geysers. Small moons have dusty rings associated (Methone, Pallene, Anthe) that may result from their erosion. More recently was discovered a very distant and huge ring, the Phoebe ring, that has a torus like structures and is composed of dust thought to be lost from the Phoebe satellite – that is on an inclined retrograde orbit – and that spiral slowly inward the Saturn's system due to the Poynting Robertson drag (Verbiscer et al., 2009). The Saturn's system shows that there is a complex interplay between moons and rings, and studies suggest that there is a kind of cycle of material in ring/satellites systems : rings may give birth to satellites at the Roche Limit, and satellite may release dusty material that moves inward due to radiation forces (see e.g. Charnoz et al., 2011, Crida and Charnoz 2012) . Is there a full recycling of material? For the moment, this is an open question (Esposito 2010).

Another intensive discussion about Saturn's rings is their age. If the rings are as old as the age of Solar system, rings would have been bombarded by micro-meteoroid, polluting ring materials through time. However, current rings look cleaner than it is expected assuming a specific pollution rate (Zhang et al. 2017, Icarus 294, 14-42) and thus rings might be young (~100 million years old). However, the flux of the micro-meteoroid bombardment through the evolution of Solar system is not well constrained and question remains how to dynamically form rings such young (see Section 3). In addition, the viscous spreading of rings of arbitrary initial mass leads in 4.5 Gyrs to rings of about half the mass of Mimas, which is the mass observed today (see below). This mass coincidence is a strong argument in favor of old rings; in this frame, the pollution could have been diluted in the initially massive rings, and stored for instance in satellites (Charnoz et al. 2011). So, the age of Saturn's rings is still an open question. Further studies and detailed data analysis of the Cassini spacecraft are required to understand more about age of Saturn's rings.

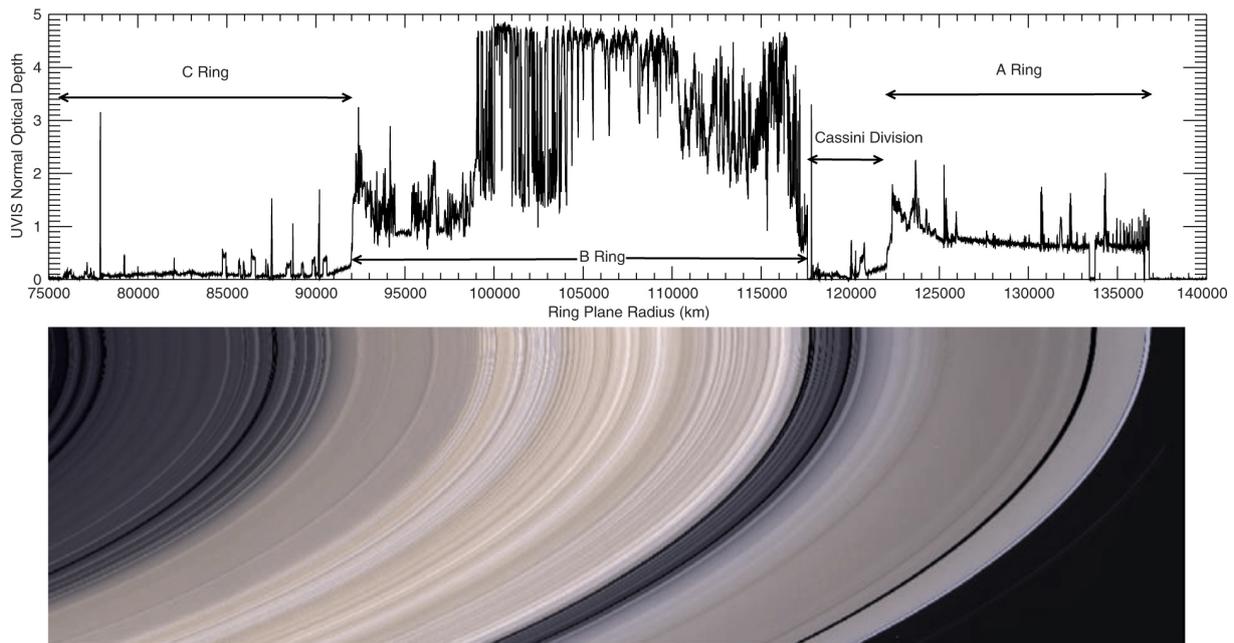

Figure 3 : Top : Saturn's main ring optical depth as a function of distance. Bottom : Saturn main rings seen in visible wavelength. Cassini ISS/JPL/NASA

**1.3 Uranus Rings**

Uranus rings are very different from those of Jupiter and Saturn. They consist mainly of a discrete collection of about 10 ringlets and two dust bands, with low albedo (< 5%). Their composition is unknown but their low albedo suggest that they are not madeof water ice (at least for a substantial part, French et al., 1991). The majority of them are not circular, but slightly eccentric and inclined, that could result from the gravitational interaction with the nearby moons, or the effect of self-gravity, that can confine the longitude of pericenters and nodes. Uranus rings are dense, made of macroscopic particles and contains little dust. Their optical depth is in range between 0.1 and 1. They have sharp edges, that may also be the result of gravitational confinement with nearby moons or the effect of self gravity.

Interestingly many moons orbit between Uranus' ringlets (for example the "Portia Group" comprises eight moons orbiting below the ν ring, between 2.3 and 4 Uranus radii). It has been proposed that meteoroid bombardment may regularly destroy some of these moonlets, which would temporarily form ringlets from the debris, that would finally reacrete into a single object (Colwell et al., 2000). So, it is thought that Uranus' ring structure may be very changing with times, and that "ring" and "moonlets" may be considered as the two-possible state of material in this region. This, again, strengthen the idea that there could be a cycle of material in planetary rings system, with exchange between moons and rings.

According to the current estimates of the Portia group mass, if it was to be fully destroyed, the resulting surface density of the ring would be about 450kg/m$^2$, comparable to Saturn's A ring (Showalter and Lissauer 2006, French and Showalter 2011, Tiscareno 2013) . Whereas the current mass of Uranus' ring is difficult to estimate it seems that there is enough mass there to make self-gravitating ring. This may have important implications for the ring system long term evolution. Particle sizes in Uranus rings range from micrometer sizes ( industy rings like the λ ring) up to cm to meter size in dense rings like the ε ring.

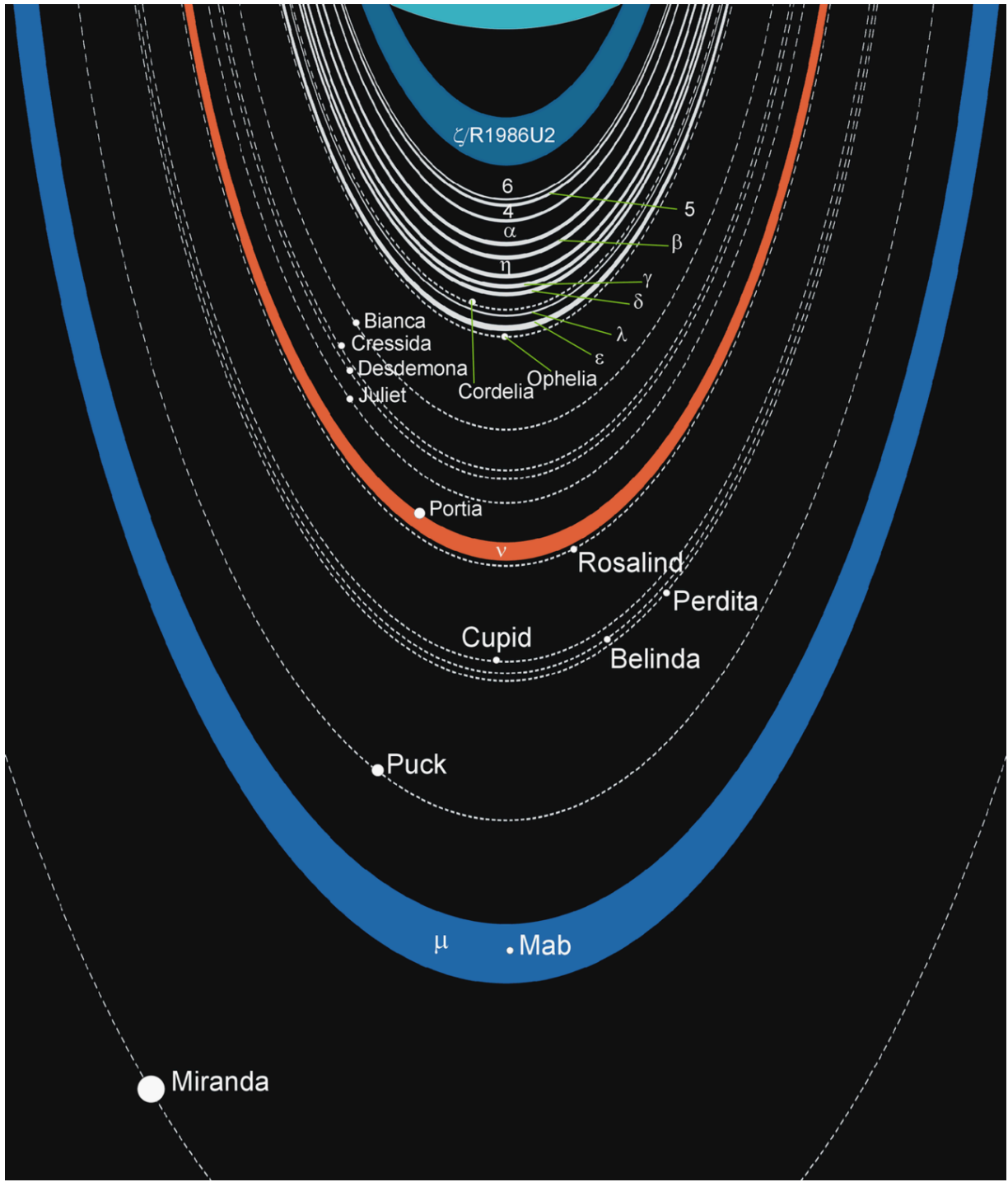

Figure 4 : Sketch of Uranus' rings and inner moons system

**1.4 Neptune Rings**

Neptune rings consists of 5 discrete rings, intertwined with numerous moonlets, recalling those of Uranus. However, they are in general dustier, made of darker material and ring's edges are much less sharp than Uranus or Saturn's one. They somewhat recall those of Jupiter because of their high dust content. One of the most notable structure of Neptune's rings is the presence of five arcs structures (i.e. incomplete rings) inside the Adam ring. It is hypothesized that these arcs correspond to material trapped in corotation sites with the moon Galatea, however this is still a matter of debate (Namouni et al., 2002). The numerous moonlets that are intertwined with the rings suggest that there is, like for Uranus' rings, material recycling between rings and moons, due to regular moon erosion or destruction because of meteoroid bombardment. The optical depth of Neptune rings ranges from $10^{-4}$ to about 0.1. Neptune rings are dominated by micrometer sized dust, like those of Jupiter.

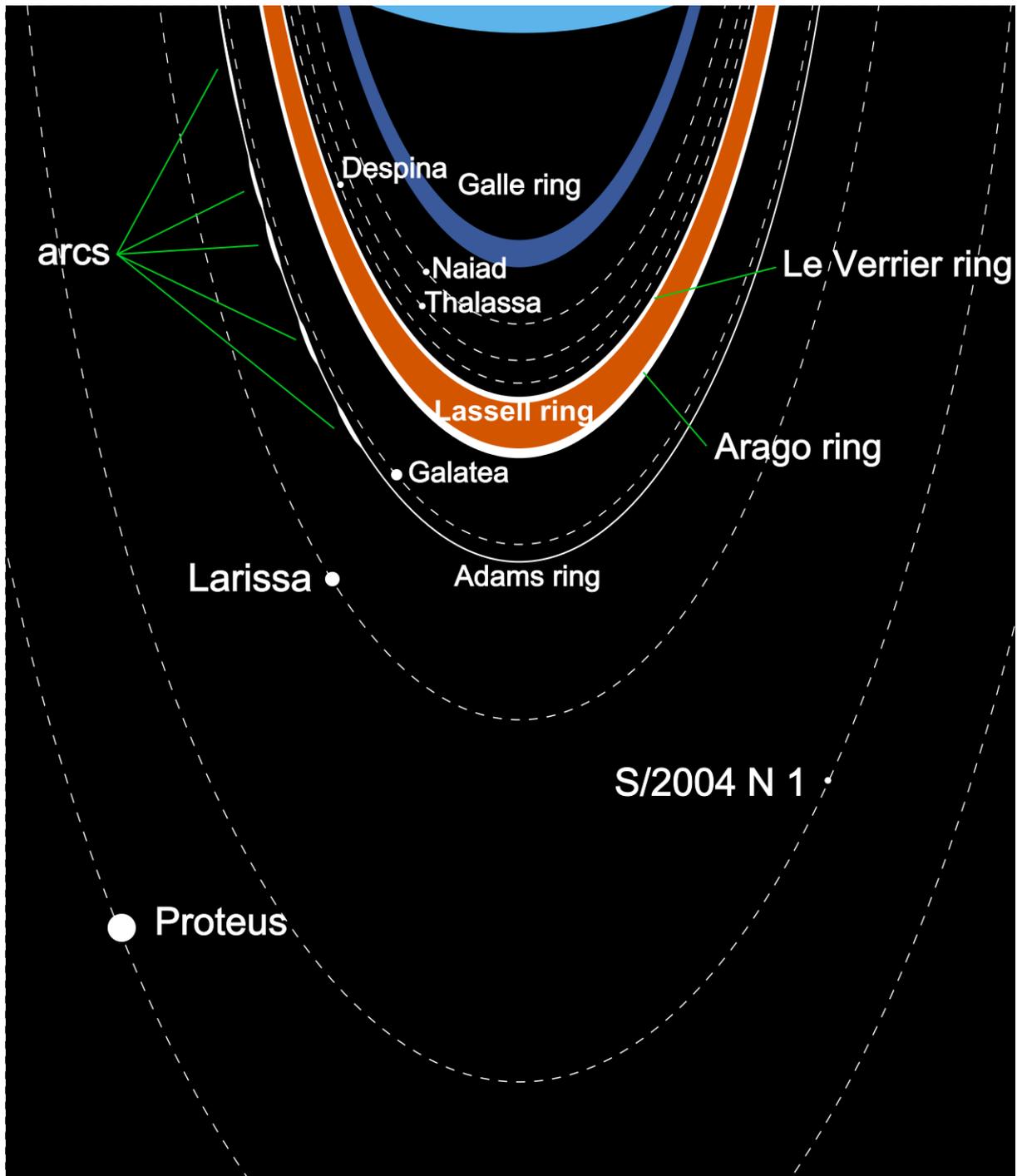

Figure 5: Sketch of Neptune's rings system and its inner moons.

## 2. Key process in rings

Rings are not static structures. They evolve through numerous, internal or external processes. Understanding their evolution is of course key to understand their origin, their frequency around (exo)planets, and even their impact on (potentially habitable) moons around giant planets in the habitable zone of their host stars. One key concept is the concept of "Roche Limit". The Roche Limit has a somewhat fuzzy definition, and slightly different definitions lead to slightly different values, whereas the concept is robust. Roche Limit, is loosely defined as the radius inside which material cannot gravitationally accrete to form large bodies and stay as individual particles. A simple derivation consists in computing the minimal distance from the planet's center for which the relative velocities between two particles on circular keplerian orbits, and with material density $\rho_m$ is larger than their escape velocity. Using these assumptions, it is found that the Roche Limit can be written

$$R_L = A\, R_P (\rho_p/\rho_m)^{1/3} \qquad \text{Eq. 1}$$

where A is a constant multiplicative factor (1.4 <A <2.5 typically), $R_P$ is the planet radius, and $\rho_p$ is the planet average density. In the example above, A=1.4, but other values are proposed in the literature: A=2.45 (Weidenschilling et al., 1984). Taking A=2.45 is often found in the literature and seems to agree well with Saturn's rings outer edge, assuming that Saturn's ring particle are made of ice. Inside the Roche limit, accretion is prevented by the tides, and rings should remains as rings. Beyond this limit, particles can accrete. Of course there is a smooth transition between these two regimes, and this limit is not sharp. For example, in Saturn's A ring, or Uranus an Neptune rings, there is a coexistence of rings and moonlets close to the Roche Limit.

## 2.1 Viscous evolution

By definition, rings are dominated by the gravity of the central body, and no pressure force needs to be considered, in contrast to the case of gaseous disks or envelopes. Hence, the orbital velocity of ring particles is dictated by Kepler's law. In the keplerian dynamics, the angular velocity and the linear velocity are decreasing functions of the orbital radius, but the specific angular momentum increases as the square root of the orbital radius. Therefore, any interaction between an inner and an outer particle that tends to lower their velocity difference results in increasing their angular momentum difference, and thus their separation. As a consequence rings spread like any astrophysical disk.

These angular momentum exchanges between the fast rotating inner and the slow rotating outer regions can be represented by a viscous stress, and if often modeled through a prescription for the equivalent viscosity $\nu$. Daisaka et al. (2001) explain that this viscosity can be decomposed in three components in dense rings of solid particles:
a) the collisional viscosity, which corresponds to actual impacts among particles
b) the translational viscosity, which corresponds to the specific angular momentum carried by particles that change orbit
c) the gravitational viscosity, which corresponds to effects of self gravity between particles.
The gravitational component is proportional to the square of the optical depth $\tau$ of the rings (or of their surface density $\Sigma$). When self-gravity is strong enough to trigger small wakes in the density distribution (as is the case in Saturn's A-ring), the gravitational component dominates, and the bulk motion associated to the transient wakes enhances the translational viscosity to a value equivalent to the gravitational one. The ring evolution is given by (e.g. Salmon et al. 2010, Eq.(3) ):

$$\partial\Sigma/\partial t = (3/r)\, \partial/\partial r[\sqrt{r}\, \partial/\partial r(\nu\Sigma\sqrt{r})] \,. \qquad \text{Eq. 2}$$

With ν given by the above prescription, the evolution of rings displays the following properties:
(i) The more massive they are, the faster self-gravitating rings evolve. This leads to the mass of the rings reaching an asymptotic evolution in $1/\sqrt{t}$ whatever their initial mass (e.g. Crida & Charnoz 2014).
(ii) The evolution of the rings slows down considerably as soon as they are not massive (or dynamically cold) enough for the self-gravity to create wakes.
(iii) As a narrow ring spreads, it should keep sharp edges, as the low density regions spread slower than the high density ones.

## 2.2 Ring/satellites interactions

For the same reason that rings spread (outward flow of angular momentum), they also repel satellites that orbit inside or outside of the rings (and reciprocally). The ring-satellite interaction theory has been developed in the 1980s following Voyager's encounter with Saturn. This interaction can be decomposed in elementary satellite-particle encounters (Lin & Papaloizou 1980) or in sum of mean motion resonances (Goldreich & Tremaine 1979), but the result is the same: the satellite should migrate away from the ring at a rate:
$d\Delta/dt = (32/27)qD\Delta^{-3}$
where $\Delta = (a-r_R)/r_R$ is the normalized ring-satellite distance, with $a$ the orbital radius of the satellite and $r_R$ the radius of the edge of the ring, $q$ is the satellite to planet mass ratio, and D in the ring to planet mass ratio.

The torque exerted by the satellite on the rings may overcome the viscous torque and stop the spreading of the rings. In Saturn's system, the 2:1 resonance with Mimas prevents the B-ring from spreading outwards into the Cassini division (see Figure 6.d), and Pan and Daphnis are able to open a gap on their orbit (Figure 6.c), by gravitationally repelling either side. But it should be noted that many thin rings around Uranus or Chariklo (a centaur) do not have known shepherds to date, and the reason of their confinement remains mysterious.

Also, small satellites (< 1km) have been detected through the wake they leave in the rings. They are called "propellers" owing to the "propeller" shape of their gravitational wake (Figure 6.a).

## 2.3 External processes : Ballistic transport and various drags

The Poynting-Robertson drag is a well-known effect in which photons emitted by the Sun perturb the orbit of dust grains. If the grains orbit around a planet, the effect of the solar photons can still be computed, and also leads to a decay of the orbital radius, in a time given by (Burns et al. 2001) :
$T =\sim 1000$ years $(a/1AU)^2(\rho/(1g/cm^3))(s/1\mu m)Q_{PR}^{-1}$        Eq. 3

where a is the semi major axis of the planet, ρ and s the density and size of the considered grain, and $Q_{PR}$ the non dimensional radiation pressure coefficient (of order 1). It clears small dust off the system.

Other drag sources can be mentioned, such as plasma drag or atmospheric drag close to the surface of a giant planet, but they are generally negligible compared to the Poynting Robertson drag.

Rings as vast and optically thick as Saturn's are an easy target for small bodies in the Solar System. Impacts with a (micro)meteorite at high velocity produce ejectas that fall back onto the rings. As a result, the angular momentum and mass are transported, leading to a decay of the ring towards the planet. This is particularly efficient in the case of low density rings such as Saturn's C ring. Also, the meteoritic bombardment could lead to ramps in density profile and waves (see Durisen et al. 1989, 1992, 1996, Charnoz et al., 2009). The bombardment by interplanetary material is also supposed to change the composition of the rings, due to the deposition of the impactor's material. In the Solar System, and in Saturn's system particularly, the rate of bombardment is unfortunately too poorly constrained to conclude of the importance of this effect in the history of Saturn's rings. Cassini should provide soon a measure of the meteoroid bombardment flux at Saturn.

**2.4 Rings as parent bodies of moons**

An interesting consequence of the ring spreading and ring-satellites interaction is the formation of moons. A gravitationally unstable media, if undisturbed, should collapse into a spherical body. The material that constitute the rings of Saturn does not, only because Saturn's tidal forces are stronger than self-gravity. However, the tidal forces decay as the distance to the planet cubed and, beyond a distance called the Roche radius, they lose to self-gravity. As rings spread beyond the Roche radius, the material coalesces into new moons, that are repelled by the rings further away, and grow by merging with one another (Charnoz et al. 2010, Crida & Charnoz 2012, see Figure 6.b). This process generates either 1 satellite if the ring spreads very fast (as was the case of the Moon-forming disk around the Earth) or a series of satellites whose mass-distance distribution follows a precise law, in excellent agreement with the regular satellites of Saturn, Uranus, and Neptune (Crida & Charnoz 2012). This suggests that this process may be universal, in the Solar System and beyond. Although the massive rings around Uranus, Neptune, and the Earth that gave birth to their satellites have now disappeared, it seems that looking for rings around exoplanets may be a way of looking for moons.

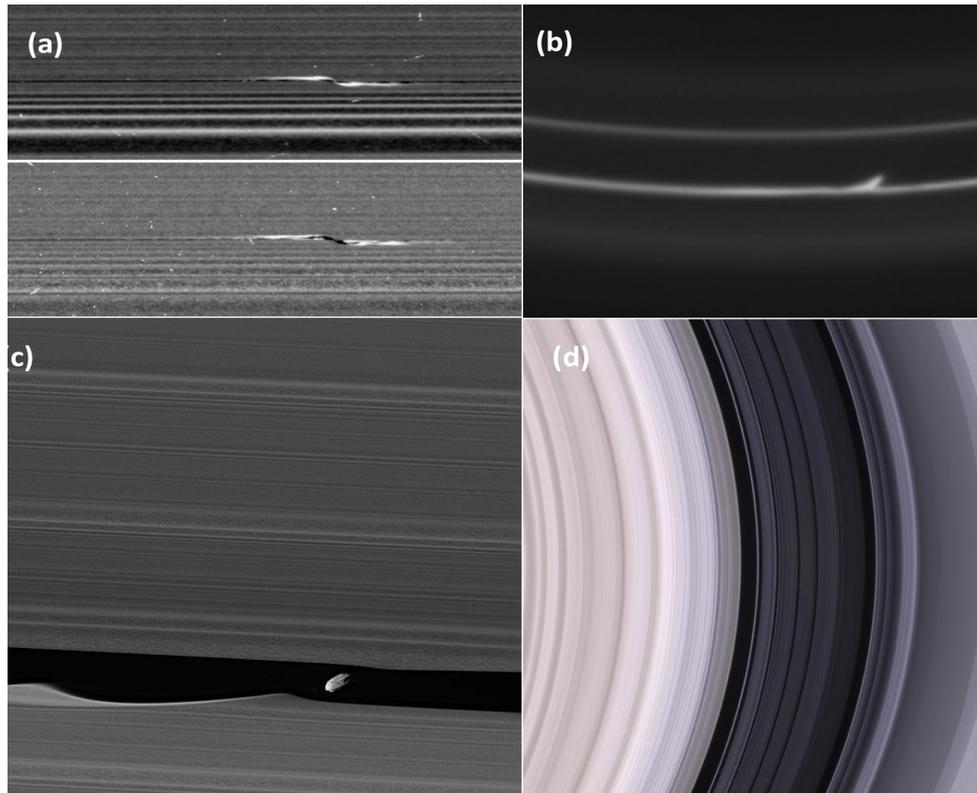

Figure 6 : Examples of dynamical structures seen in Saturn rings. (a) propellers observed in Saturn's A ring, they are the gravitational wakes produced by km sized (unseen) moons in the middle (b) ephemeral structure observed in Saturns's F ring located at Saturn's Roche Limit (c) Daphnis in Saturn's keeler division. Note the wakes produced on the edge of the Keeler gap. Above density waves are also visible in the image (d) The Cassini division. On the left side the bright B ring is visible, on the right side the translucent A ring is visible. Credits : NASA/JPL/SSI

## 3. Possible origins of planetary rings

For dusty rings systems, like Jupiter's rings systems, it seems that the erosion of the small inner moons, due to meteoroid bombardment, is able to explain the presence of the different dusty rings structures, as well as for Saturn's E, G and Iapetus rings. Uranus and Neptune present a somewhat intermediate case, where the collection of ringlets and associated moons suggests that periodic destructions of moons can generate temporary rings and reacretion of ringlets may produce temporary moons. Micrometer dust grains are easily brought inside the planet's Roche Limit, because of Poynting-Robertson drag. So the generation of dusty rings close, or inside the Roche Limit, does not pose a theoretical challenge. However, for massive rings, (like those of Saturn) or rings containing particles >> micrometer sized, how to bring material so close to the planets remains an issue.

Indeed, the origin of massive rings systems (Saturn like rings systems) around giant planets is still very debated, and for the moment no scenario seems to gather a full consensus in the researchers community.

Different scenarios have been proposed, intended mainly to the case of Saturn, to explain the origin of massive planetary rings inside the Roche Limit. Most of them imply the destruction of an ancient satellite inside, or close to, the Roche limit. But problems remain: how to bring an object so close to the planet. In Canup (2010) it is proposed that a differenciated proto-satellite migrated very close to the planet in Saturn's circumplanetary disk, posterior to its formation, and was tidally destroyed. Most of the satellite material ended inside the planet (due to migration) and the icy mantle of the satellite created a debris disk that corresponds to today rings. In this scenario the rings are as old as the planet. However, gas drag with the circumplanetary disk could have removed all debris. So there are still several caveats in this model. Another scenario proposed that Kuiper belt object, passing very close to the giant planets, could be tidally destroyed during their passage (Hyodo et al., 2017). Whereas appealing, this scenario does not explain why the four ring systems orbit in the same direction (because the geometry of encounter with a KBO shower is random). In that case the ring system could have any age, but an old ring system is preferred because the KBO flux on giant planets was higher in the past. More recently, it was proposed that collisions among satellites, due to excitation by the Sun (the so called "evection resonance") was proposed (Ćuk et al. 2016), and that the debris could spread down to the Roche Limit. However, it was shown in Hyodo and Charnoz (2017) that this process does not work and that debris re-acreation is faster than spreading. Finally, could it be possible that planetary rings are the result of the giant planet accretion process itself, and does not result from a peculiar event in a giant planet history ? This scenario has never been really investigated quantitatively (see discussion in Charnoz et al., 2009), mostly because of numerous uncertainties on the giant planets gas accretion process, however, it seems that in these dense disks the temperature and density is so high close to the planet(see e.g. Szulagyi et al. 2016) that volatile material, like water (that constitutes almost 90% of Saturn's rings mass) may not survive.

## 4. Rings around small bodies

Recent observations surprisingly revealed that rings are not only the found around giant planets but also around small objects. In this section, we briefly summarize the current observations of ringed small objects and discuss their origins.

### 5.1 Observational facts

The Centaurs are dynamically unstable minor planets that cross or have crossed the orbits of giant planets. Since the discovery of the first Centaur, 2060 Chiron (Kowal et al. 1979), a number of Centaurs have been detected. Horner et al. (2004) estimated that the total number of Centaurs with a diameter larger than 1 km is approximately ~44,000. The sources of the Centaurs are thought to be either the classical Kuiper Belt (Levison & Duncan 1997), the scattered disk (Di Sisto & Brunini 2007), or the Oort cloud (Brasser et al. 2012), and their mean dynamical lifetime is estimated to be ~$10^6$ years (Bailey & Malhotra 2009).

In 2013, a stellar occultation revealed a narrow and dense ring system around the largest Centaur 10199 Chariklo (Figure 7), the first discovery of rings around a minor planet other than the giant planets (Braga-Ribas et al. 2014). Also, similar stellar occultation observations have indicated the existence of rings around the second largest Centaur (2060) Chiron (Ortiz et al. 2015, Ruprecht et al. 2015). Interestingly, the rings found around these two Centaurs are estimated to have gaps and sharp edges, suggesting the rings are confined by shepherding satellite(s) (Braga-Ribas et al. 2014, Ortiz et al. 2015, Pan & Wu 2016). Note that, however, the rings of Chiron are still debated due to its past cometary activity that may change the interpretation of the occultation data (Meech and Belton 1989, Luu and Jewitt 1900).

In addition, not only the Centaurs but also a trans-Neptunian dwarf planets Haumea is found to possess rings around it (Ortiz et al. 2017). Haumea is one of the largest dwarf plants alongside Pluto, Ceres, Eris and Makemake and it has two satellites Hi'iaka and Namaka. The ring of Haumea is also narrow and dense. The orbital plane of the ring is almost the same as that of satellite Hi'iaka and the orbital radius of the ring is close to the 3:1 mean-motion resonance with Haumea's spin period.

So, it seems that rings might be more common than we previously thought and it is expected that more and more ringed small objects would be found among Centaurs and beyond the orbits of Neptune.

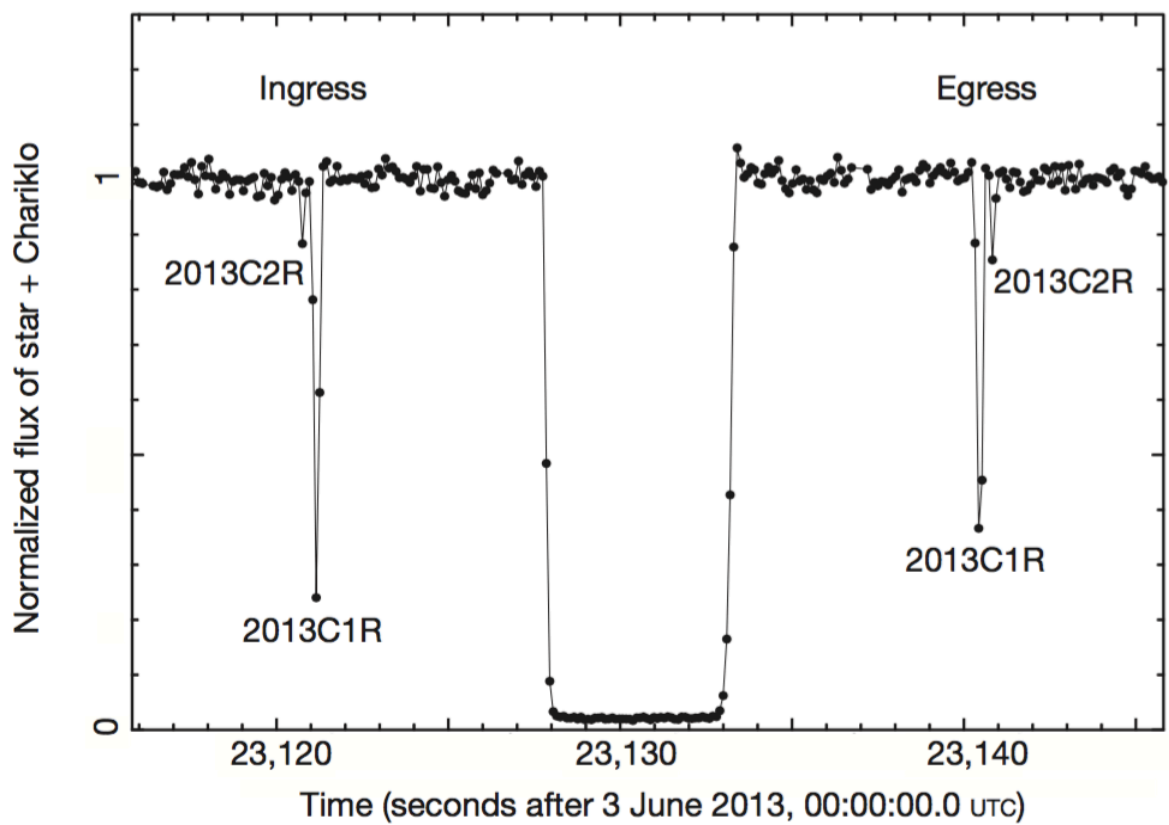

*Figure 7:* Time variation of the stellar flux observed with the Danish 1.54-m telescope (La Silla) during 3 June 2013 occultation. The central drop is due to the blocking of the stellar flux by Chariklo's main body and the two symmetric events on each side are due to the narrow rings (The figure is taken from Braga-Ribas et al. (2014)).

### 5.2 Some ideas on their origin

The dynamics and origin of rings around small objects are thought to be significantly different from those around giant planets due to their size and dynamical evolution of the parent bodies. Here, we present several possible formation scenarios for the rings around minor objects (e.g. Pan&Wu 2016, Hyodo et al. 2016): (1) ejection from the surface, (2) satellite disruption, (3) dust outgassing, (4) partial tidal disruption.

Ring formation via ejection from the surface of the object can occur through the impact or rotational disruption. Centaurs are once in the Kuiper belt and then are scattered inward. Typical collision velocity is ~1km/s in the Kuiper belt which is low-velocity collision compared to that inside asteroid belt (typically ~5km/s) and thus impact ejecta may remain as rings. However, the estimated spreading timescale of the Chariklo's rings are much shorter than the estimated collision timescale to form rings (Pan & Wu 2016). Rotational disruption of the parent body can create the rings around it. However, Chariklo's and Chiron's spin period are ~7h and the averaged spin period of the Kuiper belt objects are ~8h (Thirouin et al. 2014) which are much larger than the critical spin period for the rotational disruption. In the case of Haumea, Haumea's family and two satellites may be formed by a grazing impact (Leinhardt et al. 2010) and thus the rings may be the byproducts of the family-forming impact.

Disruption of a primordial satellite is also a possible mechanism to form rings. Such a disruption can occur via collision with a passing third object or tidal disruption as a consequence of orbital change of the primordial satellites during the close encounters with giant planet (see more details Pan & Wu 2016). In addition, as a result of the orbital change of the Centaurs from the Kuiper belt region, the mean surface temperature changes. Thus, outgassing of dust particles from the surface may occur (Pan & Wu 2016). Also, a partial disruption of a primordial differentiated object through an extreme close encounter with a giant plant may directly form debris and satellites around the remnant objects (Hyodo et al. 2016) and a long-term dynamical evolution may shape the debris to the observed narrow structures (Hyodo & Ohtsuki 2015). Either of the proposed scenario may apply to one of the observed ringed small objects, but further observations and investigations are required to constrain more about their physical properties and their origins.

## 5. Conclusion

Rings are ubiquitous to giant planets of our Solar System, but we still do not know if they are common structures among exoplanets. Discovering such structures around exoplanets, or showing that they do not exist, may certainly help to understand their origin, and in particular to decide if they are a natural byproduct of planet formation, of if they are a consequence to a subsequent evolution of the system. The forthcoming ambitious exoplanet missions, like CHEOPS, PLATO or ARIEL may help also to make further progress on our understanding of planetary rings origin. In addition, due to the close association of rings and satellites, that could be of a genetical nature or gravitational nature, discovering exo-rings, may help to infer the presence of unseen moons around exoplanets.

**Acknowledgements**
SC acknowledges the financial support of the UnivEarthS Labex program at Sorbonne Paris Cité (ANR-10-LABX- 0023 and ANR-11-IDEX-0005-02), as well as CNES funding.

**References**:

Bailey, B. L., & Malhotra, R. 2009, Icarus, 203, 155

Beurle K., Murray C.D., Williams G.A., Evans M.W., Cooper N.J., Agnor C.B., 2010. ApJ 718, L176-L180

Braga-Ribas, F., Sicardy, B., Ortiz, J. L., et al. 2014, Natur, 508, 72

Brasser, R., Schwamb, M. E., Lykawka, P. S., et al. 2012, MNRAS, 420, 3396

Burns, J. A., P. L. Lamy, and S. Soter 1979, Icarus 40, 1–48.

Burns J., Showalter M.R., Morfill G.E., 1984. In *Planetary Rings* edited by R. Greenberg and A. Brahic, pp 200-272, Univ. Arizona Press, Tucson

Burns J.; Hamilton D.; Showalter M. 2001. In Interplanetary Dust, Edited by E. Grün, B.A.S. Gustafson, S. Dermott, and H. Fechtig, p 641, Springer, Berlin.
Canup R.M., 2010. Nature 468, 943-946

Charnoz, S., L. Dones, L. W. Esposito, P. R. Estrada, and M. M. Hedman 2009. *In Saturn from Cassini-Huygens*, M. Dougherty, L. Esposito, and S. M. Krimigis (Eds.), pp. 537–575, Springer-Verlag, Dordrecht.

Charnoz S., Salmon J., & Crida, A. 2010, Nature, 465, 752-754

Charnoz S., and 12 co-authors. 2011, Icarus 216, 535-550

Colwell, J. E., Esposito, L. W., and Bundy, D. 2000. J. Geophys. Res., 105, 17589–17600.

Crida A. & Charnoz S. 2012, Science, 338, 1196-1199

Crida A. & Charnoz S. 2014, IAU#310, 182-189

Ćuk M., Dones L., Nesvorny D., 2016. ApJ 820, Id.97

Cuzzi J., R. Clark, G. Filacchione, R. French, R. Johnson, E. Marouf, and L. Spilker 2009, *in Saturn from Cassini-Huygens*, M. Dougherty, L. Esposito, and S. M. Krimigis (Eds), pp. 459–509, Springer-Verlag, Dordrecht.

Cuzzi J.F., and 22 co-authors, 2010. Science 327, 1470-1475

Daisaka H., Tanaka H., Ida S. 2001, Icarus, 154, 296-312

Di Sisto, R. P., & Brunini, A. 2007, Icarus, 190, 224

Durisen, R. H., N. L. Cramer, B. W. Murphy, J. N. Cuzzi, T. L. Mullikin, and S. E. Cederbloom. 1989, Icarus 80, 136-166

Durisen, R. H., P. W. Bode, J. N. Cuzzi, S. E. Cederbloom, and B. W. Murphy. 1992, Icarus 100, 364–393

Durisen, R. H., P. W. Bode, S. G. Dyck, J. N. Cuzzi, J. D. Dull, and J. C. White 1996 Icarus 124, 220-236

Esposito, L. W. 2010. Ann. Rev. Earth Planet. Sci., 38, 383–410

French, R. G., P. D. Nicholson, C. C. Porco, and E. A. Marouf 1991. *in Uranus*, edited by J. T. Bergstralh, E. D. Miner, and M. S. Matthews, pp. 327–409, Univ. Arizona Press, Tucson.

French, R. S., and M. R. Showalter 2011. AAS Division on Dynamical Astronomy Meeting Abstracts, 42, 6.02.

Hamilton D.P., Krüger H., 2008. Nature 453, 72-75

Horner, J., Evans, N. W., & Bailey, M. E. 2004a, MNRAS, 354, 798

Hyodo, R., Charnoz, S., Genda, H., & Ohtsuki, K. 2016, ApJL, 828, L8

Hyodo, R., Charnoz, S., Ohtsuki, K., and Genda, H. 2017. Icarus, 282, 195–213

Hyodo, R. & Charnoz, S., 2017. Astron. J. 154, 34

Hyodo, R. & Ohtsuki, K., 2015, Nature Geo. 8, 686-689

Kowal, C. T., Liller, W., & Marsden, B. G. 1979, in IAU Symp. 81, Dynamics of the Solar System (Dordrecht: Reidel), 245

Leinhardt, Z. M., Marcus, R. A., & Stewart, S. T. 2010, ApJ, 714, 1789

Levison, H. F., & Duncan, M. J. 1997, Icar, 127, 13

Lin D.N.C, Papaloizou J. 1979, MNRAS, 186, 799-812

Luu, J. X., and Jewitt, D. C. 1990. Cometary activity in 2060 Chiron. Astron. J., 100(Sept.), 913–932.

Meech, K., and Belton, M. 1989. (2060) Chiron. IAU Circ., Feb., 4770.

Namouni, F., Porco C.C 2002. Nature. 417, 45–47

Ortiz, J. L., Duffard, R., Pinilla-Alonso, N., et al. 2015, A&A, 576, 18

Ortiz, J. L., Santos-Sanz, P., Sicardy, B., et al. 2017, Nature, 550, 219

Pan, M. P., & Wu, Y. 2016, A&A, 821, 1

Reffet E., Verdier M., Ferrari C., 2015. Icarus 254, 276-286

Ruprecht, J. D., Bosh, A. S., Person, M. J., et al. 2015, Icar, 252, 271

Salmon J., Charnoz S., Crida A., Brahic A. 2010, Icarus, 209, 771-785

Schmidt, J., K. Ohtsuki, N. Rappaport, H. Salo, and F. Spahn 2009. *in Saturn from Cassini-Huygens*,

M. Dougherty, L. Esposito, and S. M. Krimigis (Eds), pp. 459–509, Springer-Verlag, Dordrecht.

Showalter M., Lissauer, J. J.; French, R. G.; et al. , 2008. American Astronomical Society. Bibcode:2008DDA....39.1602S. Retrieved 2008-05-30

Showalter, M. R., and J. J. Lissauer 2006. Science, 311, 973–977

Spitale J.N., Porco C.C. (2010) Detection of free unstable modes and massive bodies in Saturn's outer B ring. AJ 140, 1747-1757

Szulágyi, J.; Masset, F.; Lega E.; Crida A.; Morbidelli A. 2016, MNRAS, 460, 2853-2861

Tiscareno M.S., Burns J.A., Nicholson P.D., Hedman M.M., Porco C.C., 2007. Icarus 189, 14-34

Tiscareno M.S., 2013. In Planets, Stars and Stellar Systems, by Oswalt, Terry D.; French, Linda M.; Kalas, Paul, Springer Science+Business Dordrecht

Thirouin, A., Noll, K. S., Ortiz, J. L., & Morales, N. 2014, A&A, 569, A3

Verbiscer, A. J., Skrutskie M.F., Hamilton D.P., 2009. Nature, 461, 1098–1100

Weidenschilling, S. J., Chapman, C. R., Davis, D. R., and Green- berg, R. 1984. Pages 367–415 of: Greenberg, R., and Brahic, A. (eds), IAU Colloq. 75: Planetary Rings.